





 \message{Assuming 8.5 x 11 inch paper.}
 \message{' ' ' ' ' ' ' '}

\magnification=\magstep1	          
\raggedbottom
\parskip=9pt

%

\def\singlespace{\baselineskip=12pt}      
\def\sesquispace{\baselineskip=16pt}      




%



\font\openface=msbm10 at10pt
 %

\def\Minkowski     {{\hbox{\openface M}}}

 \def\dal{\displaystyle{{\hbox to 0pt{$\sqcup$\hss}}\sqcap}}



\def\lto{\mathop
        {\hbox{${\lower3.8pt\hbox{$<$}}\atop{\raise0.2pt\hbox{$\sim$}}$}}}
\def\gto{\mathop
        {\hbox{${\lower3.8pt\hbox{$>$}}\atop{\raise0.2pt\hbox{$\sim$}}$}}}
%
%
%


\def\half{{1 \over 2}}


\def\&{{\phantom a}}







\def\interior #1 {  \buildrel\circ\over  #1}     




\def\basisvector#1#2#3{
 \lower6pt\hbox{
  ${\buildrel{\displaystyle #1}\over{\scriptscriptstyle(#2)}}$}^#3}


\def\tilde{\widetilde}		


\def\&{{\phantom a}}		












\font\bmit=cmmib10			
\font\expo=cmmib10 at 10 true pt	

\newfam\boldmath

\textfont8=\bmit 
\scriptfont8=\expo 
\scriptscriptfont8=\expo

  \mathchardef\alpha="710B     \mathchardef\beta="710C
  \mathchardef\gamma="710D     \mathchardef\delta="710E
  \mathchardef\epsilon="710F   \mathchardef\zeta="7110
  \mathchardef\eta="7111       \mathchardef\theta="7112
  \mathchardef\kappa="7114     \mathchardef\lambda="7115
  \mathchardef\mu="7116        \mathchardef\nu="7117
  \mathchardef\xi="7118        \mathchardef\pi="7119
  \mathchardef\rho="711A       \mathchardef\sigma="711B
  \mathchardef\tau="711C       \mathchardef\phi="711E
  \mathchardef\omega="7121     \mathchardef\varepsilon="7122
  \mathchardef\varphi="7127    \mathchardef\imath="717B
  \mathchardef\jmath="717C     \mathchardef\ell="7160
  \mathchardef\partial"7140




\def\goesto#1{\quad\lower 1ex\overrightarrow{\ssize\hphantom{M}
    #1 \hphantom{M}}\quad}

\def\goesblank{{\;\hbox to 25pt{\rightarrowfill}\;}}





%


%
%
%
%
%
%
%
%
%
%
%
%

%
%
 \let\miguu=\footnote
 \def\footnote#1#2{{$\,$\parindent=9pt\baselineskip=13pt%
 \miguu{#1}{#2\vskip -7truept}}}
%
%

\def\linebreak{\hfil\break}
\def\lbr{\linebreak}


\def\BulletItem #1 {\item{$\bullet$}{#1 }}

\def\AbstractBegins
{
 \singlespace                                        
 \bigskip\leftskip=1.5truecm\rightskip=1.5truecm     
 \centerline{\bf Abstract}
 \smallskip
 \noindent	
 } 
\def\AbstractEnds
{
 \bigskip\leftskip=0truecm\rightskip=0truecm       
 }

\def\ReferencesBegin
{
 \raggedright			
 \singlespace					   
 \vskip 0.5truein
 \centerline           {\bf References}
 \par\nobreak
 \medskip
 \noindent
 \parindent=2pt
 \parskip=6pt			
 }

\def\section #1 {\bigskip\noindent{\headingfont #1 }\par\nobreak\noindent}

\def\subsection #1 {\medskip\noindent{\subheadfont #1 }\par\nobreak\noindent}

\def\reference{\hangindent=1pc\hangafter=1} 

\def\ref{\reference}

\def\sepref{\baselineskip=18pt \hfil\break \baselineskip=12pt}
 %

\def\journaldata#1#2#3#4{{\it #1\/}~{\bf #2 $\!$:} $\!$#3 (#4)}
 %

\def\eprint#1{{\tt #1}}
 %
 %
 %

\def\author#1 {\medskip\centerline{\it #1}\bigskip}

\def\address#1{\centerline{\it #1}\smallskip}

\def\furtheraddress#1{\centerline{\it and}\smallskip\centerline{\it #1}\smallskip}

\def\email#1{\smallskip\centerline{\it address for internet email: #1}}



\font\titlefont=cmb10 scaled\magstep2 

\font\headingfont=cmb10 at 13pt

\font\subheadfont=cmssi10 scaled\magstep1 







\def\ltilde{{\tilde l}}
\def\Rtilde{{\tilde R}}
\def\Vtilde{{\tilde V}}
\def\Mtilde{{\tilde M}}
\def\Ntilde{{\tilde N}}
\def\Atilde{{\tilde A}}
\def\kappatilde{{\tilde\kappa}}

\def\Lam{\Lambda}


\phantom{}

\def\versionnumber#1{
 \vskip -1 true in \medskip 
 \rightline{version #1} 
 \vskip 0.3 true in \bigskip \bigskip}

\versionnumber{1.5}

\sesquispace
\centerline{
  {\titlefont Big extra dimensions make $\Lambda$ too small
  }\footnote{{\titlefont *}}%
{ To appear in a special issue of the {\it Brazilian Journal of Physics}
         devoted to the Proceedings of the Second International Workshop
         ``DICE 2004'' held September, 2004, Piombino, Italy. 
	\eprint{gr-qc/0503057}}}

\bigskip


\singlespace			        

\author{Rafael D. Sorkin}
\address
 {Perimeter Institute, 31 Caroline Street North, Waterloo ON, N2L 2Y5 Canada}
\furtheraddress
 {Department of Physics, Syracuse University, Syracuse, NY 13244-1130, U.S.A.}
\email{sorkin@physics.syr.edu}

\AbstractBegins
I argue that the true quantum gravity scale cannot be much larger than
the Planck length, because if it were then the quantum gravity-induced
fluctuations in $\Lam$ would be insufficient to produce the observed
cosmic ``dark energy''.  If one accepts this argument, it rules out
scenarios of the ``large extra dimensions'' type.  
I also point out that the relation between the lower and higher
dimensional gravitational constants in a Kaluza-Klein theory is
precisely what is needed in order that a black hole's entropy admit a
consistent higher dimensional interpretation in terms of an underlying
spatio-temporal discreteness.
%
%
\AbstractEnds


\sesquispace


Probably few people anticipate that laboratory experiments which can be
done in the foreseeable future will help to guide the construction of a
theory of quantum gravity.  Even astronomers, who can access a much
greater range of conditions than we can reproduce here on earth, have
hardly provided relevant observations so far, and only in one or two
cases has any of their data been compared with the predictions of any
quantum gravity theory.  The reason for expecting this experimental
impotence to persist, of course, is that quantum gravitational effects
are normally taken to be associated with a length scale of around
$10^{-32}cm$, whereas our laboratory instruments are only able to probe
distances which are bigger than that, by fifteen or more orders of
magnitude.

But how sure are we that the quantum gravity scale is genuinely set by
the gravitational coupling constant $\kappa=8{\pi}G$ (together with
$\hbar$ and $c$, taken as unity)?  Although this is by far the most
natural assumption, no one knows with certainty that it is correct.  On
the contrary, a variety of authors have proposed in recent years that
the ``true Planck length'' is much bigger than we think it is, possibly
big enough to be accessible to the next generation of particle
colliders.  In these alternative scenarios, the $G$ that we derive
from planetary orbits and the like is not supposed to be fundamental,
because our spacetime is supposed to be only a submanifold (``brane'')
embedded in some higher dimensional spacetime, and it is the Planck length
associated with this higher dimension's gravity that would be truly
fundamental.

Were one of these modified Kaluza-Klein scenarios correct, quantum
gravity might be about to enter an experimental paradise like the one in
which particle physics thrived a few decades ago.  However, there is an
irony in the situation, as I will try to show in this paper.\footnote{$^\star$}
{In the conference, my talk was a review of causal set theory overall
 that would be difficult to reproduce here in the space allotted.  (For
 other overviews, see [1].)
 Instead, I have selected one topic from my talk (the cosmological
 constant) and developed one possible consequence of the ideas presented
 there.}
An enlargement of the Planck length sufficient to bring all these
benefits would end up depriving us of the one predictive success that
quantum gravity has had so far, namely that concerning the cosmological
constant $\Lambda$.  More specifically, what I will argue is that: if
the fluctuations [2] predicted by causal set theory give
a true account of 
the non-zero $\Lambda$ (``dark energy''), and if the fundamental
discreteness hypothesized by causal sets corresponds to the scale set by
some more fundamental, higher dimensional gravitational constant, then
making the latter much bigger than $G$ would predict fluctuations in
$\Lambda$ far too small to be compatible with its observed value.  That
is, the ``large extra dimension'' scenarios would be ruled out on this
basis.

Underlying this conclusion is the intuition that the fundamental length
posited by causal set theory, or any other discrete theory, must have an
order of magnitude given by the (true) gravitational coupling constant
$\kappa$.  Or to put it the other way around, one is supposing that the
dimensional constant $\kappa$ is more or less directly reflecting not,
for example, some length-parameter of the standard model of particle
physics, but rather a more fundamental length or ``cutoff'' in nature at
which the continuum picture breaks down.  In much the same manner ---
kinetic theory tells us --- the molecular length- and
time-scales set the order of magnitude of such continuum ``coupling
constants'' as the diffusion constant and the speed of sound.  Indeed,
one can take the point of view that without some discrete structure
underlying spacetime, there would be no good reason for the concept of
length to exist at all.
This kind of intuition or ``dimensional analysis'' is bolstered by one's
experience with renormalization in quantum field theory and in
statistical mechanics, more specifically by experience built up in
connection with the so-called ``renormalization group''.  
But perhaps its strongest support in the gravitational context comes
from our understanding of black hole thermodynamics, which almost forces
on us the idea [3]
that the entropy will ultimately be understood in terms
of some underlying discrete structures ``occupying'' roughly one unit of
horizon area each, where, in order to match the formula
$S=2\pi{A}/\kappa$, the unit of area must be that corresponding to a
length of $l\sim\sqrt{\kappa}$.

The above parameters $\kappa$ and $l$ refer specifically to four dimensions, but they
generalize immediately to dimension $D=4+d$.  There, the
(rationalized) gravitational coupling constant $\kappa$ has dimensions
of $[length]^{D-2}$ 
(in order that the term in 
the gravitational Action ${1\over2\kappa}\int RdV$ be dimensionless)
and so we may assume that 
$$
   \kappa \sim{l}^{D-2} \sim{l}^{d+2}  \ , \eqno(1)
$$
where $l$ is the fundamental length of the higher dimensional theory.


Having identified the fundamental length-scale, we must now relate it to
the cosmological constant.  Here we merely reproduce the heuristic
conclusion from causal set theory that, at any given cosmological epoch,
and in Planckian units,
$$
   \Lam  \sim{N}^{-1/2}  \ ,   \eqno(2)
$$
where $N$ is the number\footnote{$^\dagger$}
{The earlier treatments of [2] did not define $N$
 precisely.  The more complete model of [4] interprets
 $N$ ``at'' any given element $x$ as the number of elements causally
 preceding $x$ and belonging to the current cycle of cosmic expansion.
 By equation (3) this is given by the volume contained within
 the past light cone of $x$.}
of elements since the ``big bang'' [2] [4].
This long-standing prediction
correctly yielded the observed order of magnitude of the so-called ``dark
energy'' and it also explains why $\Lam$ should coincide in order of magnitude
with the density of ambient
matter.  For purposes of the present analysis, I will assume
that (2) is valid in general, even though a full comparison
of this idea with the cosmological data has not yet been completed.
In this relationship, the number $N$ is to be identified,
according to one of the basic principles of causal set theory, with the
spacetime volume in fundamental units (up to an unknown multiplicative
constant of order unity), that is:
$$
        N  \sim V / l^n       \ ,     \eqno(3)
$$
where $n$ is the spacetime dimensionality and
$$
    V = \int dV = \int d^nx |g|^\half  \ .      \eqno(4)
$$
From these equations, it is clear that changing the fundamental length
$l$ will change the predicted value of $\Lam$.  We will see that in the
scenarios under consideration, this change will lead to a current value
that is much too small.


The models in question are, more specifically, those in which the
quantum gravity energy scale is brought down to something on the order
of a $TeV$, or in any case much lower than the normal Planck scale.
Given the bemusing variety of such models, it seems difficult to make
any blanket statement covering all of them.  I will thus limit myself to
three which seem to be the sources of most of the others, namely
[5], [6] and [7].  In fact, I will limit myself primarily
to the the first of these, because the analysis is more straightforward
for it.  I believe that the conclusions would be the same for the other
two models, but I am not certain, because 
the physical interpretations of these models are
less clear to me, especially in the case of [6].

In the following, we will be comparing predictions from the usual four
dimensional theory with predictions from a hypothetical, higher
dimensional model.  To distinguish between analogous quantities in the
two cases, I will use a tilde.  Thus, for example $\kappa=\sqrt{8\pi{G}}$
will be the usual gravitational constant, while $\tilde\kappa$ will be
the analogous higher dimensional parameter, i.e. the one which is more
fundamental if the theory is true.  Similarly $l$ and $\ltilde$ will
be the corresponding fundamental length-scales.

Recall that in the original Kaluza-Klein models, spacetime was taken to
be, at least locally,\footnote{$^\flat$}
{Globally, the topology can deviate from the simple product form.  Not
 only can one have a ``twisted'' $K$-bundle over $M$, but even the local
 product structure can break down in places.  At such points the
 dimensionally reduced spacetime will be singular, betraying thereby
 the physical reality of the higher dimensions [8].}
a product manifold of the form $\Mtilde=M\times K$, where $K$ is some
compact ``internal space'', in the first instance a circle.  The
effective spacetime $M$ after coarse-graining to distance scales much
longer than the compactification diameter must thus be identified with
the {\it quotient} of the higher dimensional spacetime with respect to
$K$.  (From time to time, people would write papers in which they
identified our spacetime with a submanifold rather than a quotient, but
that was only because they didn't understand the theory!)

In these models, one sees clearly how the four and five (or higher)
dimensional gravitational constants are related.  Suppose for example that
the internal space $K$ is a circle or some other Ricci-flat manifold
like a torus.  Then the gravitational Action written in terms of the higher
dimensional metric on $\Mtilde$ will be
$$
   S_{grav} = {1 \over 2 \kappatilde} \int \Rtilde d\Vtilde
$$
while in terms of the dimensionally reduced metric on $M$ it
will be 
$$
   S_{grav} = {1 \over 2 \kappa} \int R dV
$$
Since in this case we have $R=\Rtilde$, equating the last two
expressions yields
$dV/\kappa = d\Vtilde/\kappatilde$ 
or equivalently 
$$
     \kappatilde=v\kappa
$$
because $d\Vtilde=v\,dV$,
where `$v$' represents the volume of the internal manifold $K$.
(Really we should be writing approximate rather than exact inequalities
here, because the renormalization of $\kappa$ and $\kappatilde$ has been
ignored. Indeed all quantum effects have been ignored in assuming that
the product metric is governed by the classical Einstein equations.)
Converting this relationship between $\kappa$ and $\kappatilde$ into a
relationship between $l$ and $\ltilde$ produces, in light of (1),
$$
   \left(\ltilde / l\right)^2 \; \sim \; v \, / \, \ltilde^d  \ . 
  \eqno(5)
$$
In principle, the true discreteness scale $\ltilde$ could thus be very
different from what we call the Planck length, if the compactification
diameter were very large in fundamental units.  However, in the
traditional setting, in which $M$ is a quotient of $\Mtilde$, this
would produce gauge coupling constants which were much too small.

In the more recent scenarios, though, $M$ is identified with a
4-dimensional membrane {\it within} $\Mtilde$ (a so-called 3-brane), and
the gauge fields are supposed to be confined to this membrane,
protecting them from being diluted.  It is thus that the possibility
arises of ``$TeV$ quantum gravity''.  

In order to see most simply how this possibility runs into trouble with
the cosmological constant, notice that to equate $\Lam$ to $1/\sqrt{N}$ in
natural units, is equivalently to assert that (in such units)
$$
   S_\Lam  :=  \Lam V =  \Lam N   \sim  \sqrt{N}    \eqno(6)
$$
This puts the content of (2) in truly dimensionless form,
since both $S_\Lam$ and $N$ are pure numbers.  Now if we change our
idea of the fundamental length, we will have to change $N$ as
well; and (6) teaches us that the effective cosmological constant
$\Lam$ will then change in the same ratio, or rather its square
root.\footnote{$^\star$} 
{There is a potential confusion here that needs to be avoided, between
 the cosmological constant belonging to the effective 4-dimensional
 description of the higher dimensional theory, and the true
 cosmological constant of the higher theory that would be reported by
 higher dimensional observers.  The latter actually varies inversely to
 $\sqrt{N}$ when measured in higher dimensional Planck units, but it is
 not the quantity of phenomenological interest to us.}
What is this ratio according to the scenario of reference [5]?
With the help of equations (3) and (5), we can answer this
question easily:
$$
  {\Ntilde \over N} 
  \sim
  {\Vtilde / \ltilde^{4+d} \over V / l^4}
  =
  {\Vtilde\over V} {l^4 \over \ltilde^{4+d}}
  =
  {v \over \ltilde^d} \left({l \over \ltilde}\right)^4
  \sim
  \left({\ltilde \over l}\right)^2   \left({l \over \ltilde}\right)^4
  = 
  \left({l \over \ltilde}\right)^2
$$
whence
$$
     \sqrt{\Ntilde/N}  \  \sim  \  l / \ltilde    \eqno(7)
$$
With $\ltilde$ corresponding to anything like a $TeV$, this yields a
$\Lam$ many orders of magnitude too small to do justice to the supernova
data.\footnote{$^\dagger$}
{The reasoning leading to this conclusion should transfer readily from
 the model of [5] to that of [7], with the 5-dimensional
 radius of curvature there playing the role of the compactification diameter.
 In other words $v$ here would correspond to $1/k$ there.  Notice in
 this connection that for the line-element,
 $ds_5^2=\exp(-2k|y|)ds_4^2+dy^2$, employed in [7], the volume of the
 past of an event $x$ on the brane will approximately take the same form
 as we have used herein, as is not hard to verify: 
 $\Vtilde \approx V/k \approx Vv$.  (Here, $ds_4^2$ is the line-element
 of Minkowski space $\Minkowski^4$.)
 I believe that 
 similar comments will apply as well to the model of [6], but the
 translation between four and five dimensions seems less clear
 in that model.}

In concluding, I'd like to return briefly to the basic assumption
(1) in relation to the idea that (the magnitude of) a black
hole's entropy reflects an underlying discreteness of spacetime.  Earlier,
I adduced this idea to support the identification of the discreteness
scale with the scale set by the gravitational coupling constant.  But if
this presumed relationship is consistent in four dimensions, it is not
obvious a priori that it will also be consistent with respect to the
higher dimensional spacetime.  If it were not, then one might have to
call into question, not only the ``large extra dimensions'' scenarios,
but the Kaluza-Klein paradigm in general.  Fortunately, it turns out
that Kaluza-Klein theories have no quarrel with spacetime discreteness in
this sense.  

Consider a black hole (or cosmological) horizon described with respect
to the effective, 4-dimensional spacetime $M$.  If its area is $A$, then
its entropy will be roughly $A/l^2$, $l$ being 
the fundamental length deduced from the 4-dimensional theory.  With
respect to the higher dimensional metric, however, the same horizon,
being extended over the internal manifold $K$, has a $2+d$-dimensional
``area'' of $\Atilde=Av$.  Its entropy should therefore be about
$\Atilde/\ltilde^{2+d}=Av/\ltilde^{2+d}$, if $\ltilde$ truly sets the
higher dimensional discreteness scale.  Do these two formulas agree?
Clearly, they do if and only if $(\ltilde/l)^2{\sim}v/\ltilde^d$; but
this precisely the content of equation (5).  Conversely,
equation (5) can be understood as the compatibility condition
between the lower and higher dimensional ways of ``counting horizon
molecules''.  I think that this concordance strengthens the evidence for
(1), and to that extent strengthens the evidence for the
conclusions we have drawn from it.  [If nothing else, it furnishes a
useful mnemonic for remembering equation (5)!]

In the previous paragraph, the viewpoint adopted was that of traditional
Kaluza-Klein models, for which the lower dimensional horizon is a
quotient of the higher dimensional one.  This cannot be quite correct in
scenarios such as those of [6] and [7], however.  It would be
interesting to examine the above question of ``horizon counting'' in
those cases too.


For enlightening conversations on the subject matter of this paper, I
would like to thank Yujun Chen, Jaume Gomis, Richard Corrado, and Rob Myers.

\bigskip\noindent
This research was partly supported 
by NSF grant PHY-0404646
and by funds from the Office of Research and Computing of Syracuse University.

\ReferencesBegin

\ref [1] 
Luca Bombelli, Joohan Lee, David Meyer and Rafael D.~Sorkin, 
``Spacetime as a causal set'', 
  \journaldata {Phys. Rev. Lett.} {59} {521-524} {1987}.
\sepref
Rafael D.~Sorkin,
``Causal Sets: Discrete Gravity (Notes for the Valdivia Summer School)'',
in the proceedings of the Valdivia Summer School, 
held January 2002 in Valdivia, Chile, 
edited by Andr{\'e}s Gomberoff and Don Marolf 
(to appear)\lbr
\eprint{gr-qc/0309009}.
\sepref
David D.~Reid, ``Discrete Quantum Gravity and Causal Sets''
\journaldata{Canadian Journal of Physics}{79}{1-16}{2001}
 \eprint{gr-qc/9909075}.
\sepref
Rafael D.~Sorkin,
``The causal set as the deep structure of spacetime''
 ({\it Living Reviews in Relativity}, AEI, to appear).

\ref [3] See 
  Rafael D.~Sorkin,
 ``The Statistical Mechanics of Black Hole Thermodynamics'',
  in R.M. Wald (ed.) {\it Black Holes and Relativistic Stars}, 
  (U. of Chicago Press, 1998), pp. 177-194
  \eprint{gr-qc/9705006}

\ref [2] 
Rafael D.~Sorkin, 
``First Steps with Causal Sets'', 
  in R. Cianci, R. de Ritis, M. Francaviglia, G. Marmo, C. Rubano, 
     P. Scudellaro (eds.), 
  {\it General Relativity and Gravitational Physics} 
   (Proceedings of the Ninth Italian Conference of the same name, 
     held Capri, Italy, September, 1990), pp. 68-90
  (World Scientific, Singapore, 1991).
\sepref
  Rafael D.~Sorkin,
  ``Forks in the Road, on the Way to Quantum Gravity'', talk 
     given at the conference entitled ``Directions in General Relativity'',
     held at College Park, Maryland, May, 1993,
     published in
     \journaldata{Int. J. Th. Phys.}{36}{2759--2781}{1997} \lbr
     \eprint{gr-qc/9706002}

\ref [4]  
 Maqbool Ahmed, Scott Dodelson, Patrick Greene and Rafael D.~Sorkin,
``Everpresent $\Lambda$'',
\journaldata {Phys. Rev.~D} {69} {103523} {2004}
\eprint{astro-ph/0209274}


\ref [5] 
Nima Arkani-Hamed, Savas Dimopoulos and Gia Dvali,
``The Hierarchy Problem and New Dimensions at a Millimeter'',
\journaldata{Phys. Lett.}{B429}{263-272}{1998},
\lbr
\eprint{hep-ph/9803315} .

\ref [6] 
Lisa Randall and Raman Sundrum,
``A Large Mass Hierarchy from a Small Extra Dimension''
\journaldata{Phys. Rev. Lett.}{83}{3370-3373}{1999}
\eprint{hep-ph/9905221}

\ref [7] 
Lisa Randall and Raman Sundrum,
``An Alternative to Compactification'',
\journaldata{Phys. Rev. Lett.}{83}{ 4690-4693}{1999}
\eprint{hep-th/9906064}

\ref [8]
Rafael D. Sorkin, 
``A Kaluza-Klein Monopole'',
  \journaldata{Phys. Rev. Lett.}{51}{87-90}{1983}; and {\bf 54:} 86 (1985).

\end


(prog1    'now-outlining
  (Outline 
     "\f......"
      "
      "
      "
   "\\\\message"
   "\\\\Abstrac"
   "\\\\section"
   "\\\\subsectio"
   "\\\\appendi"
   "\\\\Referen"
   "\\\\ref....."
   "\\\\end